\documentclass[apj]{emulateapj} \setkeys{Gin}{width=0.48\textwidth}
\usepackage{graphicx}

\newcommand{\hmpc}{\ifmmode{h^{-1}\,\hbox{Mpc}}\else{$h^{-1}$\thinspace Mpc}\fi}

\newcommand{\kms}{\ifmmode{\,\hbox{km\,s}^{-1}}\else {\rm\,km\,s$^{-1}$}\fi}
\newcommand{\msun}{{\rm\,M_\odot}}

\begin{document}
\title{Milky Way Halo Vibrations and Incommensurate Stream Velocities}
\shorttitle{Halo Vibrations}
\shortauthors{Carlberg}
\author{Raymond G. Carlberg}
\affil{Department of Astronomy \& Astrophysics, University of Toronto, Toronto, ON M5S 3H4, Canada} 
\email{carlberg@astro.utoronto.ca }

\begin{abstract}
Collisionless dark matter galactic halos are expected to exhibit damped oscillations as a result of ongoing late time accretion.
An n-body model of the cosmological assembly of a Milky Way-like halo is used to quantify the time
dependence of its gravitational field.
The simulation contains stellar streams with 
velocities transverse to the path of the stream are
comparable to those reported for the Orphan stream.  
The fluctuations in the quadrupole moment of the dark matter halo are measured
to be about 1\% of the radial component of the gravity field, but are 
sufficient to largely explain the perpendicular velocities. The largest late-time sub-halo within 100 kpc is
a 2\% mass at about 70 kpc from the galactic center. 
Simulations of self-interacting dark matter halos have reduced potential fluctuations as the collisions create a viscosity.
If velocity measurements of a larger sample of Milky Way streams finds (or does not find) 
the expected distribution of transverse velocities 
it will lead to limits on
the cross-section of self-interacting dark matter.

\end{abstract}


\section{INTRODUCTION}

The gravitational force from the  Milky Way's halo is conventionally and conveniently 
approximated as a static potential \citep{NFW,galpy}.
However, the late time accretion of mass and the motions of satellites means that the halo potential
is time variable at a level which can be important for some aspects of orbital motion although it does
not necessarily threaten the static approximation for other purposes.
The DR2 data release \citep{GaiaDR2:18} of the Gaia mission \citep{Gaia:16} 
has provided some intriguing new insight into the non-stationary aspects of the galactic potential.
Within a few kiloparsecs of the sun the motion of stars
 led  \citet{Carrillo:18} to reach the conclusion that the stellar distribution
around the sun is ``wobbly", with both bending and breathing modes present, with a velocity amplitude
 in the range 5-10 \kms\ kpc$^{-1}$. 
Spiral waves, the galactic bar  and bending waves 
dominate the time dependent component  potential in the region near the sun \citep{DR:19} so it is unclear 
what ongoing potential variations are present in the dark matter halo.

The stars that gravitational tides pull away from dynamically evaporating globular clusters 
and low mass dwarf galaxies create long thin streams of stars. The narrow range in
velocities of the stars means that streams are sensitive indicators of the galactic potential along the path of the stream.
The stream path, and the underlying orbits of its constituent stars,
has been extensively studied in a range
of realistic potentials with powerful methods that can be adapted to almost any frozen potential 
\citep{Dehnen:04,Choi:07,Binney:08, EB:09, SB13,Amorisco:15,BH:15,Fardal:15,BH:17}. Although systematic  offsets between the path of the stream and the orbits of its stars are expected, the offsets of the stream velocities from the path of the stream are comparable to the range that the progenitor system introduces at the outset, typically a few \kms\ for globular clusters and low mass dwarf galaxies. The stars of the Orphan stream have been found to have systematic and coherent velocities  of tens of \kms\ \citep{K19} transverse to the stream path.  That is, the velocities are incommensurate with a
fixed  stream path. The data suggest that the Milky Way halo potential must have a significant time dependence although
orbital precession in a non-spherical halo will be a contributing factor. \citet{E19} shows that a model of a static halo with the approaching
 Large Magellanic Cloud can successfully reproduce the available data.

Dark matter at the current epoch is often assumed to have a sufficiently low cross-section for interaction with
itself and other particles  that it is effectively collisionless.  
Particle dark matter at the current epoch almost certainly has
a small but finite cross-section for interaction which could help with the problem of cores of dark matter halos in dwarf
galaxies \citep{SS00}. However the past presence of gas in dwarf cores complicates their interpretation 
\citep{H1,Read19}. 
Locations which have plausibly never had much baryonic influence, such as  the halo of the Milky Way galaxy are likely
a cleaner environment to study the properties of dark matter. Self-interacting dark matter has collisions that will help reduce potential
fluctuations in a dark halo and the induced transverse velocities. 

The purpose of this paper is to assess the level of potential fluctuations that arise from the vibrations or ringing
 of a Milky Way-like dark halo due to ongoing late time mass accretion and
how that affects the orbits of stellar streams.
The approach is to use an n-body model of the build-up of a Milky Way model which contains
globular clusters which are undergoing dynamical evaporation to produce stellar streams that are directly
comparable to the density, sizes and kinematic aspects of observed streams. A set of identical n-body experiments
containing self-interacting dark matter help quantify the damping of the halo vibrations as the collisional cross-section is increased.

\begin{figure*}
\begin{center}
\includegraphics[angle=0,scale=2,trim=60 20 20 80, clip=true]{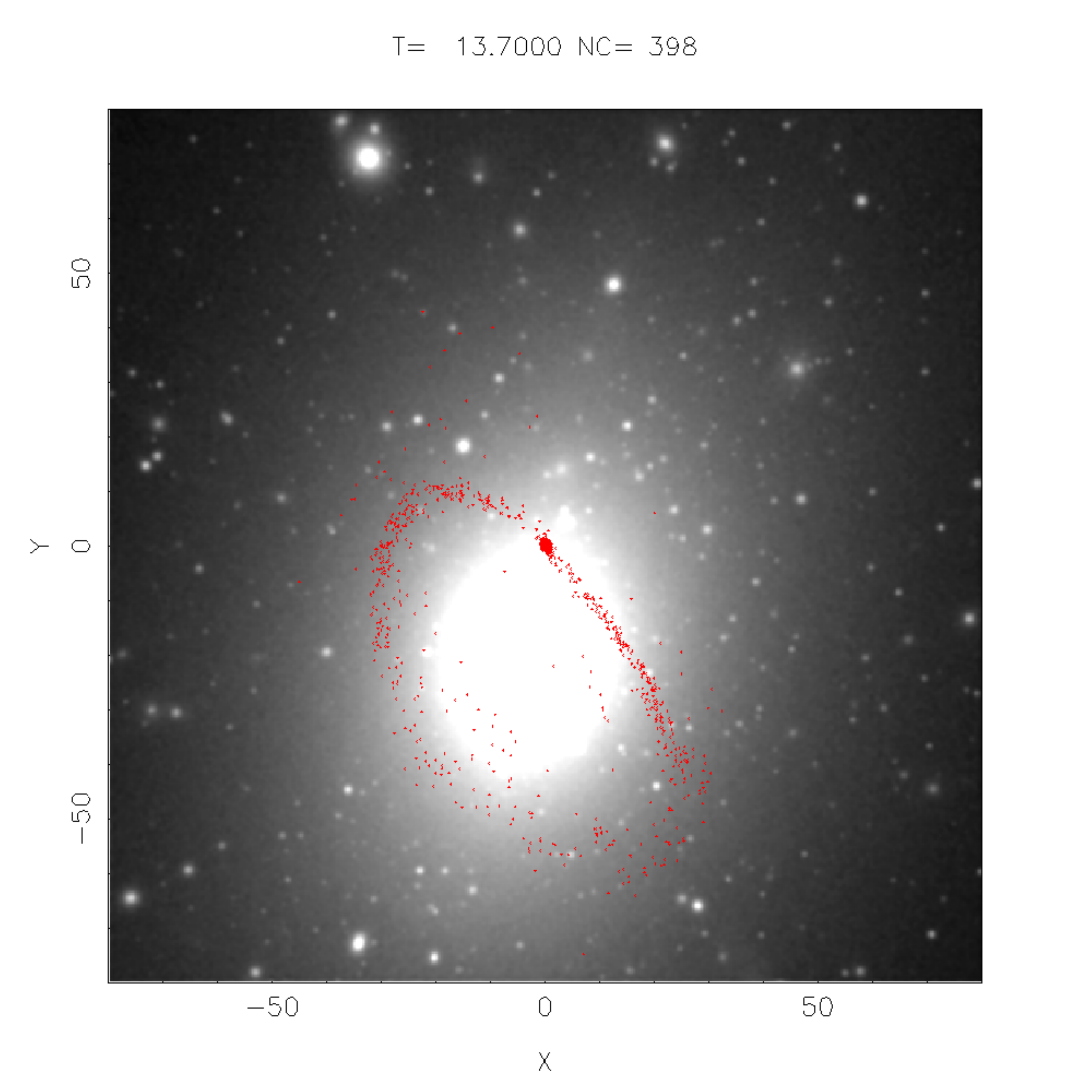}
\end{center}
\caption{The projection onto the xy plane of one set of stream particles and its progenitor (red) and the dark matter distribution (white). 
The largest sub-halo within 100 kpc is 2.2\% of the mass of the main halo and is located in the 
upper left at a distance of 68 kpc from the center.
}
\label{fig_dmstars}
\end{figure*}

\begin{figure*}
\begin{center}
\includegraphics[angle=-90,scale=1.7,trim=20 40 20 50, clip=true]{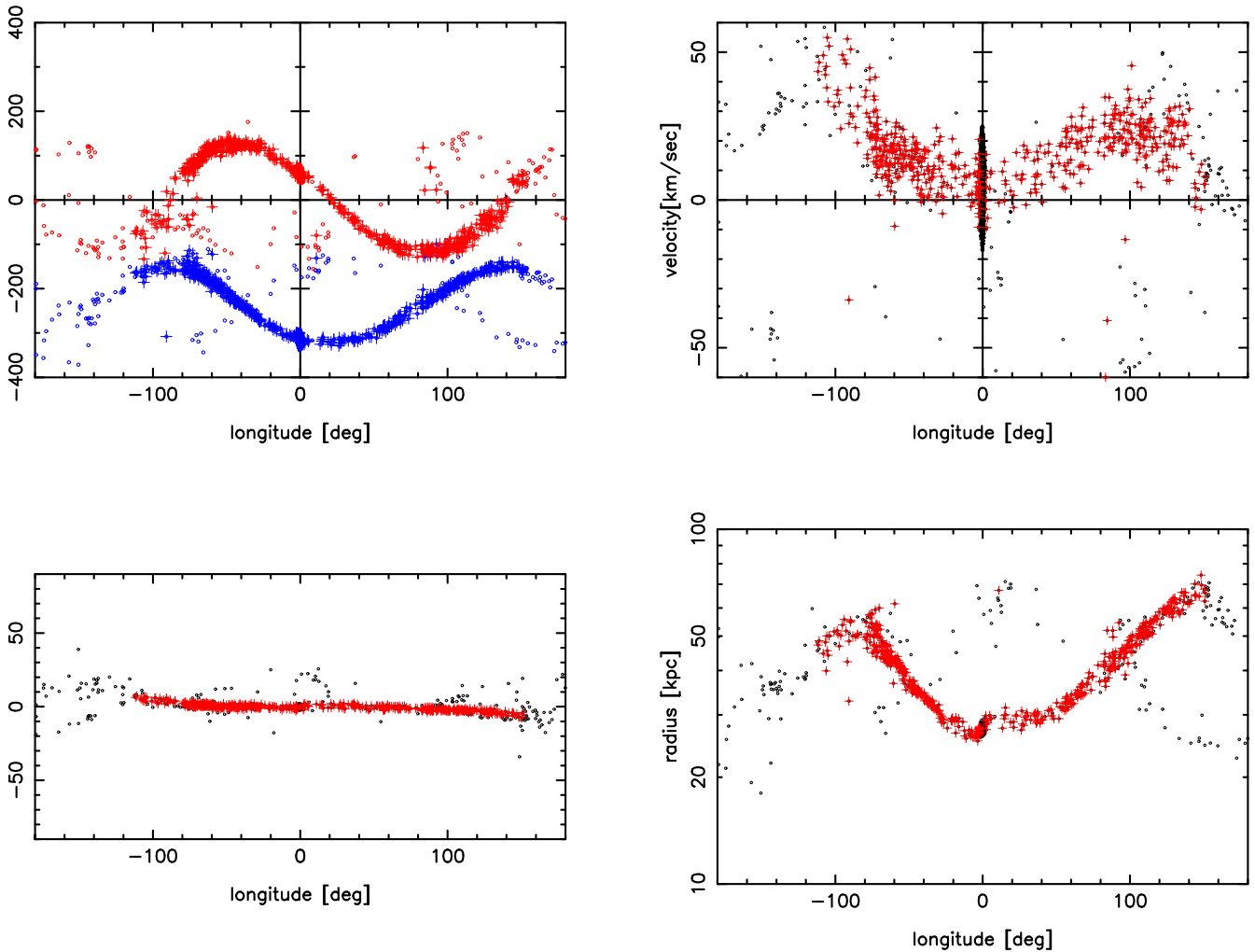}
\end{center}
\caption{The kinematics of the stream shown in Figure~\ref{fig_dmstars} as measured
from the center of the dark matter halo.
The frame is rotated to place the progenitor cluster at the origin with its velocity aligned 
to the left along the equator. 
The upper left panel shows the radial (red) and tangential velocities (blue) of
particles in the identified thin stream (+) and other cluster  particles as dots of the same color.
The upper right panel shows the velocities perpendicular to the stream,  
the lower left panel shows the stream Mercator projected on the sky,
the lower right shows the radial distance from the center, 
with thin stream as (+) and other star particles from the same cluster as black dots. 
}
\label{fig_example}
\end{figure*}

\section{Simulation with Streams}

The simulation here is the same as used in  \citet{C19} to create a dominant Milky Way-like halo from LCDM initial conditions. 
The simulation begins at redshift 4.56, reconstituting the halo catalog of the VL2 simulation \citep{VL2}. The
dark matter particles have masses of $4\times 10^4\msun$ and softening of 200 pc. 
Added to the simulation is a dynamical population of globular clusters made of 
 star particles of $5 \msun$, evolved with a softening
length of 2 pc. The simulation does not include any gas particles, or, the buildup of a central 
disk galaxy, since neither is a significant concern for the stream population beyond about 30 kpc
from the primary halo center. The outcome is a main galactic halo similar to the VL2 result with a low level of ongoing 
mass accretion. 
Pynbody \citep{pynbody} reports the instantaneous final shape of the halo in homeoidal  density shells
to have ($b/a$, $c/a$) values of  (0.96, 0.80) and  (0.94, 0.76) at 28 and 41 kpc, respectively. 
At the final moment of the simulation the Amiga Halo Finder \citep{AHF}
finds the most massive sub-halo within 100 kpc  has 2.2\% of the mass of the
$2.38\times 10^{12}\msun$ main halo  at a distance of 68 kpc from its center.
The dark matter particles and all the star particles for one stream are shown in Figure~\ref{fig_dmstars}.

\begin{figure}
\begin{center}
\includegraphics[angle=-90,scale=0.75, trim=70 20 20 50, clip=true]{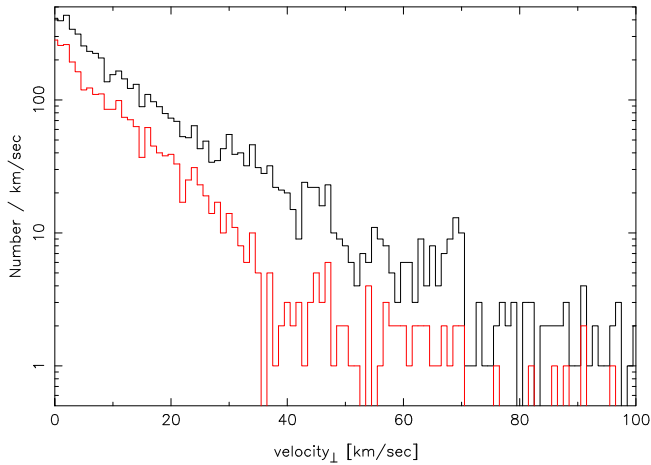}
\end{center}
\caption{The distribution of velocities perpendicular to 
the streams for all stream stars in the
range of 15-60 kpc from the center and at least 6\degr\ from their progenitor clusters.
All tidal stream particles are shown in black and the identified thin stream particles in red.
}
\label{fig_vnr}
\end{figure}

The star clusters are inserted into the simulation in the initial conditions. Their
masses are drawn from a power law cluster mass distribution over the range $2\times 10^6 \msun$
to $5\times 10^4\msun$. In total, 710  clusters are inserted into the sub-halos more massive than 
$4\times 10^8\msun$.  
The clusters lose mass as a result of internal two body relaxation and tidal fields
 to produce long, thin tidal streams that have roughly the same properties as
star streams found  in the Milky Way. The streams are identified in the distance range of 15 to 60 kpc from the center, which leads to 18
streams less than 1.2\degr\ wide on the sky. The clusters begin losing mass in their
natal dark matter sub-halo then merges into the growing dominant halo where the clusters orbit freely to produce
the thin streams.

To analyze the properties of the stream, the stars from a cluster are rotated into a frame where the progenitor cluster 
is located at the coordinate center on the sky, with the cluster moving to the left along the equator. 
Figure~\ref{fig_example} shows one stream which has properties roughly comparable  to those of the Orphan stream 
\citep{Bel:07,New:10,K19}, with the notable difference that the progenitor cluster is still present. 
This stream is one of the 18 thin streams identified in the 15 to 60 kpc range. Of those, 3  have
large and systematic velocities perpendicular to the nominal orbital plane, comparable to those in Figure~\ref{fig_example}.

Figure~\ref{fig_vnr} quantifies the distribution of  velocities perpendicular to the velocity vector of the star
cluster for stars in the distance range 15-60 kpc from the center from all streams. 
That is, all streams that qualify as thin and at least 20\degr\ long with stars 
in the selected radial range are plotted.
The  red distribution is for stars within the thin stream and the black distribution is 
for all stars both at least 6\degr\ from the progenitor clusters. The location of the two sets of stars on the sky are shown
in Figure~\ref{fig_example} in the lower left panel.  Approximating the velocity distribution in the 10-50 \kms\ range
as $\exp{(-v/v_s)}$, then $v_s$ is approximately 10  \kms\ for the thin stream.  The 
total distribution of stream stars is hard to find in imaging data but easier in kinematic data has a much 
more extended velocity distribution.
Creating the same plot as Figure~\ref{fig_vnr} at any time up to 4 Gyr earlier shows essentially the same distribution. 
That is,  significant incommensurate velocities perpendicular to the stream path on the sky are always
present at late times.

\section{Analysis of Stream Velocities}

\begin{figure}
\begin{center}
\includegraphics[angle=-90,scale=0.75, trim=70 20 20 50, clip=true]{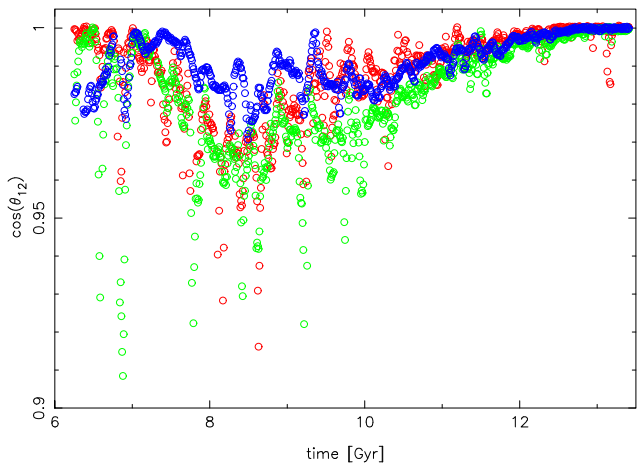}
\end{center}
\caption{The cosine of the angle between the major axis vector at the final moment and earlier times.
The circles are for measurements at 10, 20 and 40 kpc, as red, green and blue, respectively. 
The major axis tips about 10\degr\ over the last 3 Gyr.
}
\label{fig_AAmajor}
\end{figure}

The streams in the simulation have  a range of velocities perpendicular to the stream direction,
some  comparable to those in the Orphan Stream. 
The question is what aspects of the gravitational potential give rise to those velocities. The simulation does not have a 
large satellite comparable to the LMC at late times. The late time accretion is fairly modest with the mass
inside 100 kpc rising about 1\%  from 7 to 13.4 Gyr, with the mass inside 25 kpc constant with 0.7\% variance. In a 
non-spherical halo the stream will precess, leading to velocities offset from the path of the stream.
At 30 kpc a precession period of, say, 6 Gyr, leads to velocities offset from the stream,  peaking at 30 \kms,  typically about 20 \kms,
peak perpendicular to the stream.

A straightforward procedure to measure large scale fluctuations in the gravitational potential is to calculate 
the quadrupole potential as a function of radius. Because 
the vertical velocities vary slowly with longitude around the orbit, with one or two maxima in latitude, moments
higher than the quadrupole are ignored. The quadrupole potential is calculated as,
\begin{equation}
\phi_2(r^\prime) = -Gm_p\sum_{ij}  \sum_m\left[{3\over {2 r^5}} a_i a_j r_i r_j - {a^2\over{ 2r^3}}\right] .
\label{eq_qext}
\end{equation}
The $ij$  sum is over coordinate directions and the $m$ sum is over the positions of all particles of mass $m_p$.
The $m$ sum is over the $a$ coordinates for $a<r^\prime$ with $r=r^\prime$, and over $r$ coordinates 
for $r>r^\prime$, with $a=r^\prime$.
Note that only the first term generates non-radial forces.
The coordinate center is defined as the location of the highest particle density. 
To analyze the time variability of the potential the eigenvalues and eigenvectors of the quadrupole potential matrix are found.
The orientation of the largest eigenvector at the last moment (current epoch) of the simulation defines the reference direction. 
Figure~\ref{fig_AAmajor} plots the dot product of the reference with the same vector at earlier times, finding that the orientation 
of the major axis varies about 10-15\degr\ at the three radii considered. There was a significant accretion event that began around
8 Gyr, but in the last 3 Gyr the halo axis has been only slowly and systematically varying. 
The major axis of the halo wobbles, but not a lot. 
The dot product of the eigenvector associated with the third largest eigenvalue with itself shows
the orientation of the quadrupole at 25 kpc drifts at a linear rate of about 10\degr\ over the last 3 Gyr, 
but is otherwise remarkably constant.

\begin{figure}
\begin{center}
\includegraphics[angle=-90,scale=0.75, trim=70 20 20 50, clip=true]{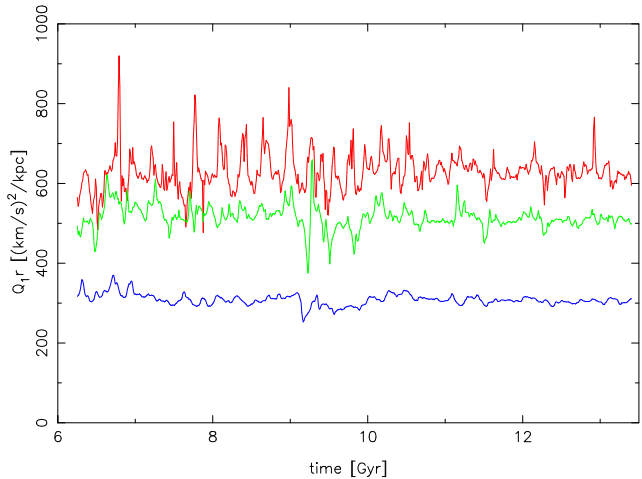}
\end{center}
\caption{The perpendicular acceleration inferred from the largest eigenvalue 
of the quadrupole potential. 
The lines are for measurements at 10, 20 and 40 kpc, as red, green and blue, respectively. 
The mean level reflects the approximately constant in time ellipsoidal shape. 
}
\label{fig_Et}
\end{figure}

\begin{figure}
\begin{center}
\includegraphics[angle=-90,scale=0.75, trim=70 20 20 50, clip=true]{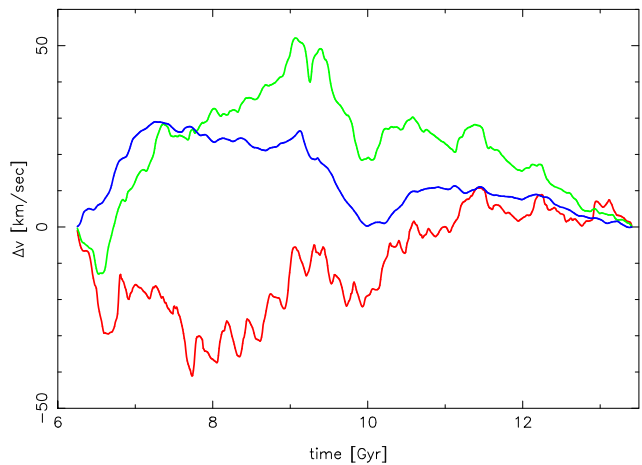}
\end{center}
\caption{The variation in perpendicular velocity that the quadrupole moment can induce with time,
calculated as the time integral of the accelerations in Figure~\ref{fig_Et} via Equation~\ref{eq_dv}.
The lines are for measurements at 10, 20 and 40 kpc, as red, green and blue, respectively. 
The calculation is done with a zero mean force over the interval.
}
\label{fig_vint}
\end{figure}

The gravitational acceleration monopole has a mean of $2300 (\kms)^2$/kpc with a variance of $17 (\kms)^2$/kpc over the
9.2 to 13.4 Gyr interval at a radius 25 kpc.  The monopole 
acceleration is radial and cannot create transverse velocities, although once present,
non-radial velocities will vary as stars with the radial excursions of an orbit.

At fixed $r$ the maximum perpendicular acceleration is 
$a_z= {3 \over 2} Q_1 r$, where $Q_1$ is the largest eigenvalue of the quadrupole. 
Values of $Q_1 r$ (without the ${3\over2}$) are plotted in 
Figure~\ref{fig_Et}. At 25 kpc the quadrupole term $Q_1 r$ has a mean and variance of $448$ and $24(\kms)^2$/kpc, 
respectively, with a range from
$305-538 (\kms)^2$/kpc over the 7.3 Gyr interval analyzed.
To estimate the scale of the velocities that can be induced perpendicular to an orbit the
time variable component of the acceleration normal to the plane is integrated over time,
\begin{equation}
\Delta v (t,r)= {3\over 2}r\int_{t_0}^t [Q_1(t^\prime,r) -\langle Q_1(r) \rangle] \, dt^\prime,
\label{eq_dv}
\end{equation}
where $\langle Q_1(r) \rangle$ is the average of the eigenvalue over the time interval. 
The fluctuating velocity change integral, Equation~\ref{eq_dv}, leads to the values shown in
Figure~\ref{fig_vint},  demonstrating that the time variations of the quadrupole potential 
can create transverse velocities of 10-20 \kms. Such velocities
would rise and fall with inward and outward excursions of the orbit.
The dipole term has a mean and variance of $23$ and $14 (\kms)^2$/kpc, respectively and is not dynamically
very important, as expected deep in a halo.

The conclusion is that the semi-coherent fluctuations of the quadrupole leads to accelerations that produce velocities sufficient to cause 
the observed perpendicular velocities of the streams.  The variation can be characterized as a vibration or a strongly damped ringing oscillation.

\section{Self-Interacting Dark Matter}

\begin{figure}
\begin{center}
\includegraphics[angle=-90,scale=0.75, trim=70 20 20 50, clip=true]{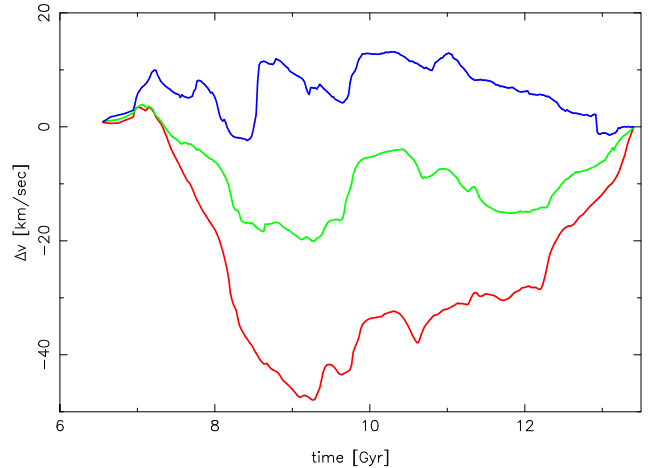}
\end{center}
\caption{The same indicative calculation of the amplitude of the perpendicular velocities as shown 
in Figure~\ref{fig_vint}, now for the velocities that 
can be expected at 25 kpc for halos composed of self-interacting dark matter particles 
with cross-sections of 0.1, 0.3 and 1.0  ${\rm cm}^2 {\rm gm}^{-1}$, shown in
red, green and blue, respectively. The calculation sets the initial and final velocity to zero.
}
\label{fig_vintSIDM}
\end{figure}

Halo ringing is damped through the collisionless mixing of the accreted dark matter into the halo of the galaxy although significant
substructure continues to circulate. If the dark matter self-interacts with a cross-section that leads 
to a mean free path, $\ell$, in the range of tens of kiloparsecs, then kinetic theory gives the kinematic viscosity as 
$\nu\simeq \sigma_v \ell$, where $\sigma_v$ is the one-dimensional velocity dispersion of the dark matter.
 Viscosity transports momentum and will increase damping, potentially to a level that would largely suppress the
halo vibrations. Self-interacting dark matter (SIDM) simulations 
can quantify the importance of the effect. 

Figure~\ref{fig_vintSIDM} shows the significance of the effect for a SIDM model with elastic collisions, as implemented 
in the Gadget3 code for a VL2 setup as before but with dark matter only and started at redshift 3. 
The cross-sections used are 0.1, 0.3 and 1.0 in units of ${\rm cm}^2 {\rm gm}^{-1}$. 
At 0.1 ${\rm cm}^2 {\rm gm}^{-1}$ the result is almost the same as collisionless dark matter. The
effect of reduced force fluctuation amplitudes and hence reduced velocities is very clear, with the fluctuations becoming very much smaller than
collisionless dark matter at a cross-section of 1.0.  If dark matter did have this cross-section then the observed perpendicular velocities
of the Orphan stream would be in fairly good accord with the static halo LMC model \citep{E19}.
A simulation at 
a cross-section of 10 units is so heavily damped that the primary effects are to cause the halo to become nearly spherical with a substantial core.

If the Milky Way halo does not have the level of halo ringing expected for
ongoing late time accretion it would provide a limit on the size of the dark matter self-interaction cross-section.
Streams will provide increasingly good measurements on the shape of the halo, with current results finding that the disk-corrected shape of the halo in the stream
region is spherical within about 5\% \citep{Bovy:16}. Of course baryons and the development of a stellar disk also effect the shape of the halo 
\citep{Bryan:13,Butsky:16,Chua:19} and the two together are even more complex \citep{Despali:19}. In general the effects diminish outwards as
the orbital frequencies of the halo become lower than the regions where baryons are present reducing dynamical coupling. Gas and stars will also 
be able to help damp halo ringing although this is not yet quantified.

\section{Discussion and Conclusions}

The Orphan Stream is observed to have velocities incommensurate with the path of the stream. 
That is, there are significant transverse velocities. The implication is that the streams are in a time varying halo potential. 
This paper shows
that ongoing late time accretion of dark matter sub-halos into the Milky Way-like galactic halo of the VL2 simulation 
leads to quadrupole potential fluctuations which can induce the transverse velocities in the 10-20\kms\ range 
that the star streams in the simulation are measured to have. 
One of the 18 streams present in the simulation is roughly at the distances of the Orphan stream and has density, length and 
roughly comparable transverse velocities. In this case the largest sub-halo within 100 kpc  is 2.2\% of the mass of the main halo 
and at 68 kpc. 

Self-interacting dark matter damps halo vibrations. A set of SIDM simulations shows that the ringing is strongly
suppressed for elastic SIDM at a cross-section of 1${\rm cm}^2 {\rm gm}^{-1}$ and that at 10 ${\rm cm}^2 {\rm gm}^{-1}$
the halo is forced to become nearly spherical. If it is shown that the velocities of other streams in the Milky Way halo 
are consistent with a near-spherical, static, halo and the LMC alone, it would provide indirect support for SIDM, although a more
accurate overall model would need to account for the baryonic damping as well.

\acknowledgements

Positive criticism from an anonymous referee led to substantial improvements in the paper.
This research was supported by  NSERC of Canada. Computations were performed on the niagara supercomputer at the SciNet HPC Consortium. SciNet is funded by: the Canada Foundation for Innovation; the Government of Ontario; Ontario Research Fund - Research Excellence; and the University of Toronto.

\end{document}